\UseRawInputEncoding
\documentclass[conference]{IEEEtran}
\IEEEoverridecommandlockouts
\usepackage{cite}
\usepackage{amsmath,amssymb,amsfonts}
\usepackage{algorithmic}
\usepackage{graphicx}
\usepackage{textcomp}
\usepackage{xcolor}
\usepackage[left=1.62cm,right=1.62cm,top=1.8cm]{geometry}
\usepackage{balance}
\def\BibTeX{{\rm B\kern-.05em{\sc i\kern-.025em b}\kern-.08em T\kern-.1667em\lower.7ex\hbox{E}\kern-.125emX}}
\begin{document}

\title{Self-assess Momentary Mood in Mobile Devices: a Case Study with Mature Female Participants\\}

\author{\IEEEauthorblockN{1\textsuperscript{st} Caterina Senette}
\IEEEauthorblockA{\textit{IIT} \\
\textit{National Research Council}\\
Pisa, Italy \\
caterina.senette@iit.cnr.it}
\and
\IEEEauthorblockN{2\textsuperscript{nd} Maria Claudia Buzzi}
\IEEEauthorblockA{\textit{IIT} \\
\textit{National Research Council}\\
Pisa, Italy \\
claudia.buzzi@iit.cnr.it}
\and
\IEEEauthorblockN{3\textsuperscript{rd} Maria Teresa Paratore}
\IEEEauthorblockA{\textit{ISTI} \\
\textit{National Research Council}\\
Pisa, Italy \\
maria.paratore@isti.cnr.it}
}

\IEEEoverridecommandlockouts
\IEEEpubid{\makebox[\columnwidth]{978-1-6654-5357-8/23/\$31.00~\copyright2023 IEEE\hfill} \hspace{\columnsep}\makebox[\columnwidth]{ }}
\maketitle

\IEEEpubidadjcol

\begin{abstract}
 Starting from the assumption that mood has a central role in domain-specific persuasion systems for well-being, the main goal of this study was to investigate the feasibility and acceptability of single-input methods to assess momentary mood as a medium for further interventions in health-related mobile Apps destined for mature women. To this aim, we designed a very simple android App providing four user interfaces, each one showing one interactive widget to self-assess mood. Two widgets report a hint about the momentary mood they represent; the last two do not have the hints but were previously refined through questionnaires administered to 63 women (age 45-65) in order to reduce their expressive ambiguity. Next, fifteen women (age 45-65 years) were recruited to use the app for 15 days. Participants were polled about their mood four times a day and data were saved in a remote database. Moreover, users were asked to fill out a preliminary questionnaire, at the first access to the app, and a feedback questionnaire at the end of the testing period. Results appear to prove the feasibility and acceptability of this approach to self-assess momentary mood in the target population and provides some potential input methods to be used in this context.
\end{abstract}

\begin{IEEEkeywords}
momentary mood, mobile devices, well-being Apps, design, input methods
\end{IEEEkeywords}

\section{Introduction}
Recent advances in technology have resulted in a growing interest in emotion and computers since a wide range of emotions undeniably plays a critical role in every computer-related and goal-directed activity. Focusing on ICT (Information and Communication Technology) systems based on goal-directed activities, capturing emotional factors is crucial to guiding the user toward his/her goals.
Among these ICT systems, we are interested in tools for health-related behavior interventions that promote behavioral changes to improve a person's health while reducing potential health risk factors. Specifically, we are interested in self-care approaches released through information systems, to manage the menopause transition and thus reduce the associated cardio-metabolic risk \cite{senette2018persuasive, lee2015understanding}. In 2014, a study by Spruijt-Metz et al. \cite{spruijt2014dynamic} advanced the framework just-in-time adaptive interventions (JITAI) which aims to \textit{“provide information, nudges, and interventions when needed or appropriate, tailored to the individual's needs and context, via mobile technologies”} \cite{spruijt2014dynamic}.  This type of framework is a promising reference methodology for developing smartphone health and fitness Apps, as well as solutions for other ubiquitous and pervasive computing technologies exploiting persuasion strategies. Starting from the assumption that mood has a central role in domain-specific persuasion systems for well-being, since it helps in recognizing \textit{when and how} to deliver behavioral interventions, the main goal of the current work is to investigate the feasibility and acceptability of single-input methods to self-assess the user's momentary mood in a specific target population. Literature recognizes that mature people, often have ambivalent attitudes towards technology and tend to be reluctant to accept technological innovations. Moreover, they often have longer learning times and lower performance measures than younger people \cite{czaja2006factors, holzinger2007some}. Thus, focused studies, like the current one, are necessary since we expect that the interface design would affect the acceptability and feasibility of the usage of momentary mood trackers in this target population. 

Momentary mood is defined as \textit{“The psychological state identified by perceived internal psychological and physical sensations, such as dullness on waking or exhilaration after walking”} \cite{sakairi2013development}. Investigating the feasibility and acceptability of tools assessing momentary mood could be helpful not only in checking user availability for mobile just-in-time interventions \cite{spruijt2014dynamic} but also in other contexts such as the Ecological Momentary Assessment (EMA) and Experience Sampling Method (ESM) \cite{shiffman2008ecological, trull2009using}, antecedent's analysis of addictive behaviors and mood regulation per se. To this aim, we designed a very simple android App providing four user interfaces, each one showing one interactive widget to self-assess mood. Two widgets report a hint about the momentary mood they represent; the last two do not have the hints but were previously refined through a questionnaire administered to 63 women (age 45-65) in order to reduce their expressive ambiguity. Next, fifteen women (age 45-65) were recruited to use the App for 15 days. Participants were polled about their mood four times a day and data were saved in a remote database. Moreover, users were asked to fill out a preliminary questionnaire, at the first access to the App, and a feedback questionnaire at the end of the testing period. Thus, main research questions were: (RQ1) \textit{Is the proposed approach for mood assessment feasible and accepted in this target population?} (RQ2) \textit{Are there differences among the proposed input methods in terms of input speed, intuitiveness, expressiveness, appropriateness to the scope, and  willingness to reuse? If so, do these differences have statistical significance?}
\section{Related work}
Several studies in the literature investigate the use of methodologies that require participants to self-report on their activities, emotions, or other elements of their
daily life several times a day (EMA/ESM). Among them, we are interested in research studies reporting mood assessments through mobile devices. A valuable literature review covering contributions from the computer science domain can be found in the work of Van Berkel et al. \cite{van2017experience}. Here the authors focused especially on mobile devices, providing important considerations that were helpful for our work, especially those related to data quality in data collection. The data collection was also the focus of another interesting study by Chan et al. \cite{chan2018students} that investigated participants’ experiences in EMA studies, providing helpful insights on how these experiences may influence the collected data. Both studies guided especially our design choices. It is worth noting that the majority of studies were conducted in a medical context and participants were people with specific health conditions. In the work of Richmond et al. \cite{richmond2015feasibility} participants were invited to respond to SMS text messages sent out weekly over a period of 15 consecutive weeks, asking participants to rate their experience of depression on a simple 9-point scale. Other studies have shown the feasibility of using EMA to gather data on bipolar disorder symptoms via smartphone \cite{bauer2005does,schwartz2016daily}, sometimes asking patients to respond to auto-generated surveys twice a day \cite{schwartz2016daily}. All these studies provide some evidence for the utility of daily mood reporting via modern software tools although additional evidence is needed to better evaluate the usability and acceptability of these tools especially among different user groups not necessarily medicalized. Focusing on methods for assessing momentary mood, Desmet et al. \cite{desmet2016mood} concentrated their attention on tools requiring minimal effort from users who have little time or poor motivation to report their moods. To this aim, they proposed and validated Pick-A-Mood, a character-based pictorial scale for reporting and expressing moods. The research closely related to our investigation is a study by Wallbaum et al. \cite{wallbaum2016comparison}, in which the authors explored the use of four different methods to express mood on mobile devices through a comparative study evaluating intuitiveness, inconvenience, speed of input, everyday use, expressiveness and overall suitability of these methods. Nevertheless, our study is different because: (i) mood assessment does not serve as a part of interpersonal communication; (ii) the target population includes only females aged 45-65 years; (iii) tools tested are different (apart from the PAM, see Section \ref{sec:pamTool}) since the study dates back to 2016 when mobile user interface design and interaction were different from the current ones.
\section{Theoretical Background}
\subsection{How to define Mood}
Psychological theories of affect highlight the importance of discriminating between emotion and mood. These terms are often confused, not only in everyday thinking but sometimes by experts as well. This is why some of the existing tools to assess affect (questionnaires, emotion-tracking Apps, and physiological sensors) are all used to track emotion as well as mood. Nevertheless, instruments for measuring emotions are not always suitable for measuring mood \cite{desmet2016mood}. Summarizing, main differences are: (i) \textit{Object-directness}, emotion is generally intentional (external focus), mood is non-intentional (internal focus) \cite{frijda1994varieties};
(ii) \textit{Time-duration}, emotions tend to be relatively short-lived (seconds, minutes), are episodic and generally express strong intensity. Moods, in contrast, have a long duration (hours or days), they are continuous and result in a week  \cite{watson1988development}; (iii) \textit{Effects}, emotions bias action, stimulating an immediate response, on the contrary, mood influences people without interrupting their ongoing behavior and thoughts, modifying general readiness to engage \cite{brave2007emotion}.

\subsection{Reference Models}
Among the main psychological theories of affect, we are interested in dimensional theories \cite{russell1999core} that focused on a mood state -- in particular, a momentary psychological condition resulting from psychological and physical sensations (such as exhilaration after walking). The dimensional model describes emotions in two (sometimes three) independent dimensions (arousal, pleasure) in a Cartesian Space. Based on these two dimensions, Russell created a \textit{Circumplex model of emotions} \cite{russell1980circumplex} most commonly used to test stimuli of emotion words, emotional facial expressions, and affective states. Every language has its mood-related words, and since our study took place in Italy, final terms, chosen from the Immediate Mood scale \cite{nahum2017immediate} and ITAMS, Italian Mood Scale \cite{quartiroli2017development} are: (i) Red-quarter - nervosa, tesa, preoccupata, arrabbiata; (ii) Yellow-quarter - attenta, energica, felice, euforica; (iii) Green-quarter - soddisfatta, serena, rilassata, calma; (iv) Blu-quarter - apatica, triste, depressa, stanca. 
\section{Design study}
To conduct our study, we recruited 15 participants (female, age range 45-65 years, M = 55.67 years, SD = 5.40 years) from women working on our research campus (about 1500 workers), for a 15-day study. Participants were independent of the study and the research team, and their involvement was voluntary. Since we are interested in health prevention paths, the age group should be temporally consistent with 'preventive' interventions. All of them were asked to use their personal device and to download our Android app available via Google Play. The app provided four user interfaces, each showing one interactive widget to self-assess mood. Participants were polled about their mood four times a day. The time of assessment could be modified with respect to the default plan (10:00/14:00/18:00/21:00) keeping a period of at least 3 h from each time. At each assessment time, the app proposed a randomly selected widget, without repetition in order to propose all four tools to each participant every day. When the participant first accessed the app, a preliminary questionnaire (pre-questionnaire) was proposed, to gather the user's information in terms of time available for oneself, presence/absence of pathological conditions, presence/absence of mood disorders and familiarity with the use of mobile Apps in general and specifically mood tracking Apps. All these data could be useful for explaining user behavior (ex post). At the end of the 15 days, each participant was asked to answer a post-questionnaire looking for subjective impressions of the app's use (see Section \ref{sec:user_perception} and Section \ref{sec:discussion}).
\subsection{Tools}
\begin{figure*}
 \centering
\dimen0=\dimexpr(\textwidth-6\tabcolsep)/5\relax
\begin{tabular}{*4{p{3.2cm}}}
\includegraphics[width=3.2cm]{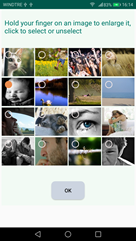} &
\includegraphics[width=3.2cm]{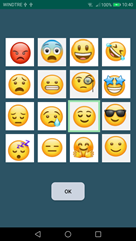} &
\includegraphics[width=3.2cm]{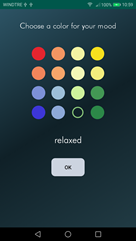} &
\includegraphics[width=3.2cm]{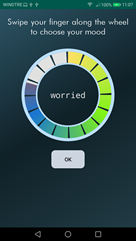} \cr
\hskip1.5em$(a)$ & \hskip1.5em$(b)$ & \hskip1.em$(c)$ & \hskip1.em$(d)$ 
\end{tabular}
\caption{Designed tools: (a) PAM widget; (b) Emoji widget; (c) Color picker widget; (d) Color wheel widget.}\label{fig:tools}
\label{fig:fig2}
\end{figure*}

Two of the four widgets derive from the literature and were appropriately adapted to our specific context, while the other two are relatively new proposals. All of them are described in the following and shown in Fig. \ref{fig:fig2} (a), (b), (c), (d).
\subsubsection{PAM widget - Photographic Affect Meter}
\label{sec:pamTool}
Pollak et al. presented the Photographic Affect Meter (PAM) in 2011 \cite{pollak2011pam} based on the Affect Grid by Russell et al. \cite{russell1989affect} and validated through PANAS \cite{watson1988development}. The PAM shows 16 photographs arranged in a 4x4 grid used to represent individual current mood. Since the final data set of images has not been shared, researchers who want to use the PAM tool must build their own set of images. To this aim, we collected images from the Italian Flickr repository, tagged by millions of users. For each word selected to express Italian terms for moods (see paragraph 3.2), we extracted a group of 5-6 pictures. Further refinement was obtained by sharing an online survey (through the mailing list and Facebook) asking people to select the most appropriate picture for each mood term. We collected answers from 125 people, 29 males and 96 females, aged between 18 and 73 years. Of these, we selected data from women aged 45-65 years (n = 63) to better represent our specific target group. Among these data, we found sufficient polarized responses (low ambiguity) for nine mood terms, for which one of the proposed options was chosen in at least 60\% of the cases, the remaining 40\% (or less) being dispersed among the other images (options) proposed for each term. For the remaining mood terms (7) we selected the most voted image, having a minimum of \%30 of choices (higher residual ambiguity). The final image set will be available to other researchers if requested.
\subsubsection{Emoji Widget}
According to the same dimensional model, we created another self-report tool based on commonly known emoji, structurally and stylistically richer than emoticons to better express mood nuances. To build our 16-item emoji tool while reducing its ambiguity as much as possible, we shared an online survey asking people to select the most appropriate emoji for each mood term, choosing from a set of 3-6 items (redundant). We collected answers from 125 people aged between 18 and 73 years (31 males and 94 females) and from them, we selected data from women aged 45-65 years (n = 63). In this case, polarized responses were higher than in the image selection (PAM widget), 12 mood terms matching with the same emoji in 60\% of the sample. For the remaining four mood terms (as occurred for the PAM widget), we selected the most voted emoji having a minimum of 30\% of choices (less residual ambiguity).
\subsubsection{Color Picker Widget}
The third tool is a variant of the Mood Meter developed by the Yale Center for Emotional Intelligence.  As the inventor stated, it helps people develop skills such as recognizing, understanding, labeling, expressing, and regulating emotions. The Mood Meter is a square evenly divided into four color quadrants. Each color quadrant represents a mood category according to Russell's model \cite{russell1980circumplex}. The color strategy in itself does not contain information about mood; the color is only a way to identify and label it. In the original tool, each quadrant has 25 emotion words with a total of 100; in our version, we selected four words for each quadrant in order to guarantee the same expressiveness and adequate usability on small-screen devices for all the widgets.
\subsubsection{Color Wheel Widget}
The fourth component is a sort of mix between the Circumplex model \cite{russell1980circumplex} and Plutchik's Wheel of Emotions \cite{plutchik2003emotions}. As with the Color Picker widget, Plutchik's wheel of emotions helps us to enhance emotional literacy by providing words for emotions, as well as to understand how different emotions are related to one another and how they tend to change over time. To be consistent with the other widgets and to guarantee better usability on smartphones, we created a simplified version of the tool, where only two levels of emotions (16 mood terms) have been exploited in a single circle. Moreover, labels associated with each element of the wheel are placed on the component center. Compared to the original tool, we preserved concepts such as the intensity level of emotions, the fusion of primary emotions into new emotions, and the concept of emotional opposites as mutually exclusive couples.
\section{Results}
\subsection{User sample characterization}
Results from  the preliminary questionnaire show that 46\% of women had difficulty finding time for themselves, 76\% were in good health; 30\% had some mood disorder but do not usually track mood. Moreover, most of the sample (70\%) do not use health Apps or tracking mood Apps. Most of them (67\%) regularly use other generic Apps, especially social networks.
\subsection{Mood assessment data}
\begin{figure*}
 \centering
\dimen0=\dimexpr(\textwidth-6\tabcolsep)/5\relax
\begin{tabular}{*2{p{7.2cm}}}
\includegraphics[width=8.2cm]{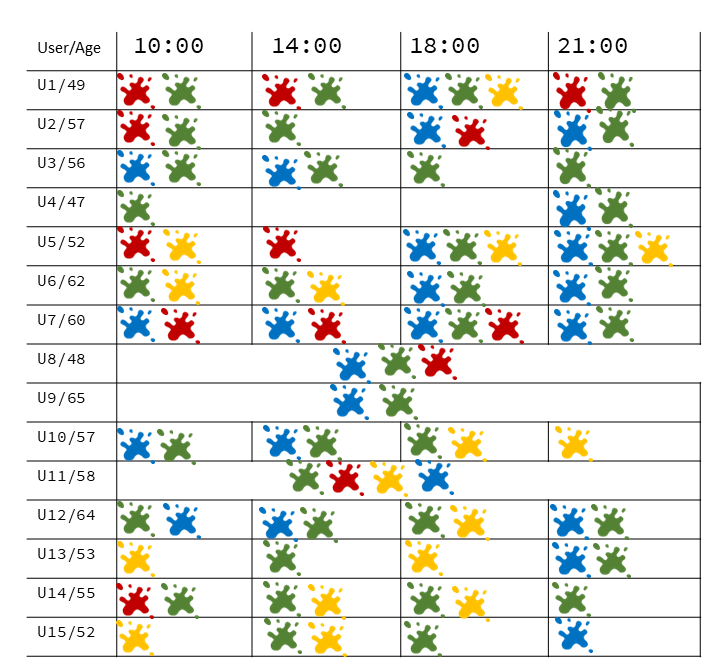} &
\includegraphics[width=6.2cm]{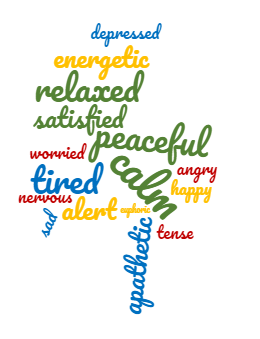} \cr
\end{tabular}
\caption{Dominant Russell's quarters during the day for each user (left); Tag cloud of selected mood terms (right).}\label{fig:tools}
\label{fig:fig3}
\end{figure*}

The analysis of mood assessment data showed good engagement by most of the users, and nearly all the participants entered the value of their mood four times a day as requested. As shown in the literature, notifications on mobile devices can cause user annoyance due to the interruption of other activities, especially ongoing tasks on the mobile device itself \cite{horvitz2001notification,van2017experience}. Determining the time of assessments is crucial to gaining user attention, but it is very difficult to achieve. For these reasons, we let the user customize the notification time according to their habits. Twelve women did not customize time-setting notifications, and the remaining three did so partially or completely. Moreover, for those who did not modify the default time setting, a time stamp associated with each recorded mood showed that 34\% of total assessments were made within 5 min from the notification event. User reactivity to notifications was generally high: the average delay (min) of responses to notifications was in the range of 5-25 min for 9 participants out of 15, and 5-40 min for the remaining 6. As suggested in the work of Pejovic and Musolesi \cite{pejovic2014interruptme}, the willingness to reply increases when a user has not received notifications for a while. Probably our target group is less engaged in continuous activities on the smartphone, and they suffer less from interruption load than other types of users. Considering the users' behavior during the day, we observed high variability among users and there was not a moment of the day more/less likely to cause assessment delays. Analyzing recorded values, we did not observe a recurrent mood term in a certain widget nor recurrences in mood values at a specific time during the day. Nevertheless, we observed that 50\% of users always selected the mood within two Russell's quarters regardless of the widget (for widgets with quarters) in each assessment time. Red-quarter's moods (energized/unpleasant feelings) and Yellow-quarter's moods (energized/pleasant feelings) are more frequent earlier in the day. Quiet feelings (pleasant or unpleasant) are more frequent in the evening. In Fig. \ref{fig:fig3} (left), we show the dominant Russell's quarters for each time assessment and for each user. We call \textit{dominant} a quarter that received a number of selections greater than the random threshold.
As shown in In Fig. \ref{fig:fig3}, users U8, U9, and U11 customized the time assessment thus, for them we cannot evaluate dominant Russell's quarters at the time specified. Moreover, other than preferred Russell's quarters, data also revealed a prevalence in the terms chosen: \textit{Calm} had the most choices (18\% of total assessments), the least frequently chosen terms were \textit{Euphoric} (o.65\%) and \textit{Angry} (1.63\%). The terms in the middle were \textit{Tired} with 12\% of choices, and \textit{Peaceful} with 13\%. Fig.\ref{fig:fig3} (right) shows a tag cloud highlighting frequencies in mood term choices as weights.
\subsection{User perception}
\label{sec:user_perception}
Other data were collected via a post-questionnaire including three sections. In the first section we asked participants to evaluate each widget in terms of \textit{input speed, intuitiveness, expressiveness, appropriateness to the scope, and willingness to reuse} (via 5-item Likert scale: 1 - strongly disagree, 5 - strongly agree). Statements were formulated in positive form, e.g.: “Assessing mood is rapid with this widget”. Results obtained by summing the scores for each widget and for each of the five variables under study are described in Table \ref{tab:friedmanTest} in terms of Best Widget (BW) and Worst Widget (WW). The Color Picker reached the best score for three of the five variables. The Emoji was perceived as the worst widget for all the variables except \textit{Willingness to reuse}.

The second section asked participants to rate each widget from 1 (poor) to 5 (excellent) regardless of specific criteria. Summing the scores collected for each widget we obtained the following order from best to worst: 1) Wheel; 2) Color Picker; 3) Emoji; 4) PAM. The third section allowed participants to leave comments and suggestions regarding each widget. Only 4 out of 15 women left comments mainly claiming the perceived ambiguity of PAM and Emoji widgets as well as the attractiveness of PAM. 
\begin{table*}[]
 \centering
 \caption{Results from user perception in terms of Best widget (BW) and Worst Widget (WW) for the five variables. Statistical report in dark grey: Friedman Test results via Chi-Square index and p-values for the same 5 variables; k is the number of interfaces (input methods), N is the user's sample dimension and DoF represents the experiment degrees of freedom.}
\scalebox{1.2}{\begin{tabular}{ cc }

\begin{tabular}[t]{l|lllcl}
\hline
\textbf{Variable} & \textbf{BW} &\textbf{Score}  & \textbf{WW} &\textbf{Score} \\ \hline
\textit{Input speed}        & Color Picker & 64 & Emoji & 36\\
\textit{Intuitiveness}      & Color Picker & 60 & Emoji& 39  \\
\textit{Expressiveness}      & Wheel & 56 & Emoji & 41\\
\textit{Appropriateness}     & PAM & 59 & Emoji & 46 \\
\textit{Willingness to reuse} & Color Picker/PAM & 51 & Wheel & 46  \\
\hline
\end{tabular}
\begin{tabular}[t]{|lllcl}
\hline
\color{darkgray}{\textbf{k}} & \color{darkgray}{\textbf{N}} & \color{darkgray}{\textbf{Chi-Square}} & \multicolumn{1}{l}{\color{darkgray}{\textbf{DoF}}} & \color{darkgray}{\textbf{p-value}} \\ \hline
\color{darkgray}{4} & \color{darkgray}{15} & \color{darkgray}{\textbf{12.14}} & \color{darkgray}{3} & \color{darkgray}{0.007} \\
\color{darkgray}{4} & \color{darkgray}{15} & \color{darkgray}{\textbf{13.96}} & \color{darkgray}{3} & \color{darkgray}{0.003} \\
\color{darkgray}{4} & \color{darkgray}{15} & \color{darkgray}{\textbf{14.42}} & \color{darkgray}{3} & \color{darkgray}{0.002} \\
\color{darkgray}{4} & \color{darkgray}{15} & \color{darkgray}{\textbf{5.44}}  & \color{darkgray}{3} & \color{darkgray}{0.142} \\
\color{darkgray}{4} & \color{darkgray}{15} & \color{darkgray}{\textbf{9.75}}  & \color{darkgray}{3} & \color{darkgray}{0.021} \\ \hline
\end{tabular}
	\label{tab:friedmanTest}
\end{tabular}}
\end {table*}
\subsection{Statistical Analysis}
The next step was to understand whether the differences identified among the input methods (via post-questionnaire) were statistically significant or they were due to chance (RQ2). The statistical perspective must consider two limits in our study, the smallness of the sample and the data provenance from Likert scales. Moreover, since each one of the four interfaces has been tested by the same group of 15 users, we are in the case of \textit{within-groups experiments}. All these things considered, the candidate tests are non-parametric tests that do not require making assumptions about the sampled population. Among those, we applied the Friedman test \cite{friedman1937use} with the Chi-Square index, adequately modified to take into account the peculiarity of the Likert scale in the rank matrices.
Results, shown in Table \ref{tab:friedmanTest}, suggest that the differences found in the four interfaces, according to each of the five dependent variables can be considered statistically significant with p-value $<$ 0.05 for all the variables except the \textit{Appropriateness} for which is not possible to refuse the null hypothesis. The Table \ref{tab:friedmanTest} also indicates how the strength of significance is lower for the variable \textit{Willingness to reuse} with a Friedman test value slightly higher than the critical value, while it is significantly higher for the other variables considered. It is worth noting that, due to the test adopted, the rejection of the null hypothesis (for each variable) guarantees that at least one pair of interfaces among the four considered, express statistically significant differences. Additional Wilcoxon Sign-Rank tests \cite{wilcoxon1950some} helped us identify which interface pairs have statistically significant differences for each variable: (i) \textit{Input speed}, all pairs have statistically significant differences except Color Picker/Color Wheel and Emoji/PAM; (ii) \textit{Intuitiveness}, all pairs have statistically significant differences except Color Picker/Emoji and Emoji/PAM; (iii) \textit{Expressiveness}, all pairs have statistically significant differences except Emoji/PAM; (iv) \textit{Willingness to reuse}, all pairs have statistically significant differences except Color Picker/PAM and Emoji/PAM.
\section{Discussion}
\label{sec:discussion}
Collected data produced ambivalent results. About 50\% of participants were willing to reuse all the widgets and 25\% of the sample took a neutral position, moreover, considering appropriateness for the scope, all the widgets were judged appropriate; both results partially indicate the success of the approach per se, regardless of the widget. Moreover, to capture "the moment", the variable to be privileged is the \textit{input speed} and this is higher for Color Picker and Wheel which are textually enriched. Nevertheless, we need to consider the trade-off between input speed and attractiveness which could affect user motivation. All the widgets force the users to search for their own mood and this search could be less attractive for textual labels. PAM widget has been judged the most appropriate and, based on the comments from participants, photos, despite being ambiguous, seem to have a potential for attractiveness. However, in the case of PAM, it is unclear to what extent the photo itself affects the mood rather than faithfully reporting it. The woman could select the mood she would like to be in rather than the one she is actually in. Arbitrary responses dues to widget ambiguity as well as poor user motivation are responsible for low data accuracy which is the main limitation of single-input methods. On the other hand, a ground truth would require standardized tools (mainly questionnaires including several items) which are impossible to map on a single-input method. The differences that emerged among the four input methods are statistically significant although the analysis has its major limitations in the smallness of the sample and in the subjectivity of the data collected. Further tests involving a larger number of users in the same population are needed to assess the actual acceptability of this approach, as the general literature reports a high dropout rate in this type of study \cite{chan2018students}. Furthermore, to avoid poor data quality due to arbitrary responses, the overall design could be enriched through incentives or rewards, focusing on time as a key factor. The look-screen chosen requires that the answers be provided in a short time because our goal is to capture the moment; in other contexts, micro-interaction would not be an advantage and interface prompts should encourage weighted responses depending on the nature of the assessment.
\section{Conclusion}
The goal of this study was to evaluate the feasibility and acceptability of using single-input methods for self-assessing momentary mood in mobile devices, in a very specific target population (female, age range 45-65). Considering that almost 50\% of the women participants declared they were usually busy during the day and not inclined to use Apps for mood tracking, test participation was positive. The app is quite minimal; it does not disturb but also does not offer anything in return, so this makes user retention riskier. Nevertheless, no participants left the test before 15 days. 
Regarding the input methods proposed, each one showed positive and negative aspects, and improving their design is certainly possible, for instance by better selecting color nuances for Color Wheel and Color Picker. While for the PAM widget, better visibility of the photos was ensured (in our PAM widget all photos could be enlarged for better visibility but some users did not use this feature) since our target population (age 45-65) is more likely to have vision problems on a mobile screen. Furthermore, to increase the attractiveness, more photos (user's choice) could be proposed (at each assessment) in case those proposed were not considered exhaustive; this could reduce intrinsic ambiguity without forcing all the widgets to be associated with textual labels. In summary, the experience described seems to prove the feasibility and acceptability of this approach to self-assess momentary mood in the target population. These results could be of help for designing richer and more attractive Apps, destined to mature women, that monitor the user's well-being and embed a module for momentary mood assessment.
\bibliographystyle{IEEEtrans}
\bibliography{mood}

\begin{thebibliography}{10}
\providecommand{\url}[1]{#1}
\csname url@samestyle\endcsname
\providecommand{\newblock}{\relax}
\providecommand{\bibinfo}[2]{#2}
\providecommand{\BIBentrySTDinterwordspacing}{\spaceskip=0pt\relax}
\providecommand{\BIBentryALTinterwordstretchfactor}{4}
\providecommand{\BIBentryALTinterwordspacing}{\spaceskip=\fontdimen2\font plus
\BIBentryALTinterwordstretchfactor\fontdimen3\font minus
  \fontdimen4\font\relax}
\providecommand{\BIBforeignlanguage}[2]{{%
\expandafter\ifx\csname l@#1\endcsname\relax
\typeout{** WARNING: IEEEtranS.bst: No hyphenation pattern has been}%
\typeout{** loaded for the language `#1'. Using the pattern for}%
\typeout{** the default language instead.}%
\else
\language=\csname l@#1\endcsname
\fi
#2}}
\providecommand{\BIBdecl}{\relax}
\BIBdecl

\bibitem{bauer2005does}
M.~Bauer, N.~Rasgon, P.~Grof, L.~Gyulai, T.~Glenn, and P.~C. Whybrow, ``Does
  the use of an automated tool for self-reporting mood by patients with bipolar
  disorder bias the collected data?'' \emph{Medscape General Medicine}, vol.~7,
  no.~3, p.~21, 2005.

\bibitem{brave2007emotion}
S.~Brave and C.~Nass, ``Emotion in human-computer interaction,'' in \emph{The
  human-computer interaction handbook}.\hskip 1em plus 0.5em minus 0.4em\relax
  CRC Press, 2007, pp. 103--118.

\bibitem{chan2018students}
L.~Chan, V.~D. Swain, C.~Kelley, K.~de~Barbaro, G.~D. Abowd, and L.~Wilcox,
  ``Students' experiences with ecological momentary assessment tools to report
  on emotional well-being,'' \emph{Proceedings of the ACM on Interactive,
  Mobile, Wearable and Ubiquitous Technologies}, vol.~2, no.~1, pp. 1--20,
  2018.

\bibitem{czaja2006factors}
S.~J. Czaja, N.~Charness, A.~D. Fisk, C.~Hertzog, S.~N. Nair, W.~A. Rogers, and
  J.~Sharit, ``Factors predicting the use of technology: findings from the
  center for research and education on aging and technology enhancement
  (create).'' \emph{Psychology and aging}, vol.~21, no.~2, p. 333, 2006.

\bibitem{desmet2016mood}
P.~M. Desmet, M.~H. Vastenburg, and N.~Romero, ``Mood measurement with
  pick-a-mood: review of current methods and design of a pictorial self-report
  scale,'' \emph{Journal of Design Research}, vol.~14, no.~3, pp. 241--279,
  2016.

\bibitem{friedman1937use}
M.~Friedman, ``The use of ranks to avoid the assumption of normality implicit
  in the analysis of variance,'' \emph{Journal of the american statistical
  association}, vol.~32, no. 200, pp. 675--701, 1937.

\bibitem{frijda1994varieties}
N.~H. Frijda \emph{et~al.}, ``Varieties of affect: Emotions and episodes,
  moods, and sentiments.'' 1994.

\bibitem{holzinger2007some}
A.~Holzinger, G.~Searle, and A.~Nischelwitzer, ``On some aspects of improving
  mobile applications for the elderly,'' in \emph{International Conference on
  Universal Access in Human-Computer Interaction}.\hskip 1em plus 0.5em minus
  0.4em\relax Springer, 2007, pp. 923--932.

\bibitem{horvitz2001notification}
E.~C. M. C.~E. Horvitz, ``Notification, disruption, and memory: Effects of
  messaging interruptions on memory and performance,'' in \emph{Human-Computer
  Interaction: INTERACT}, vol.~1, 2001, p. 263.

\bibitem{lee2015understanding}
M.~Lee, B.-c. Koo, H.-s. Jeong, J.~Park, J.~Cho, and J.-d. Cho, ``Understanding
  women's needs in menopause for development of mhealth,'' in \emph{proceedings
  of the 2015 Workshop on Pervasive Wireless Healthcare}, 2015, pp. 51--56.

\bibitem{nahum2017immediate}
M.~Nahum, T.~M. Van~Vleet, V.~S. Sohal, J.~J. Mirzabekov, V.~R. Rao, D.~L.
  Wallace, M.~B. Lee, H.~Dawes, A.~Stark-Inbar, J.~T. Jordan \emph{et~al.},
  ``Immediate mood scaler: tracking symptoms of depression and anxiety using a
  novel mobile mood scale,'' \emph{JMIR mHealth and uHealth}, vol.~5, no.~4, p.
  e6544, 2017.

\bibitem{pejovic2014interruptme}
V.~Pejovic and M.~Musolesi, ``Interruptme: designing intelligent prompting
  mechanisms for pervasive applications,'' in \emph{Proceedings of the 2014 ACM
  International Joint Conference on Pervasive and Ubiquitous Computing}, 2014,
  pp. 897--908.

\bibitem{plutchik2003emotions}
R.~Plutchik, \emph{Emotions and life: Perspectives from psychology, biology,
  and evolution.}\hskip 1em plus 0.5em minus 0.4em\relax American Psychological
  Association, 2003.

\bibitem{pollak2011pam}
J.~P. Pollak, P.~Adams, and G.~Gay, ``Pam: a photographic affect meter for
  frequent, in situ measurement of affect,'' in \emph{Proceedings of the SIGCHI
  conference on Human factors in computing systems}, 2011, pp. 725--734.

\bibitem{quartiroli2017development}
A.~Quartiroli, P.~C. Terry, and G.~J. Fogarty, ``Development and initial
  validation of the italian mood scale (itams) for use in sport and exercise
  contexts,'' \emph{Frontiers in psychology}, vol.~8, p. 1483, 2017.

\bibitem{richmond2015feasibility}
S.~J. Richmond, A.~Keding, M.~Hover, R.~Gabe, B.~Cross, D.~Torgerson, and
  H.~MacPherson, ``Feasibility, acceptability and validity of sms text
  messaging for measuring change in depression during a randomised controlled
  trial,'' \emph{BMC psychiatry}, vol.~15, no.~1, pp. 1--13, 2015.

\bibitem{russell1980circumplex}
J.~A. Russell, ``A circumplex model of affect.'' \emph{Journal of personality
  and social psychology}, vol.~39, no.~6, p. 1161, 1980.

\bibitem{russell1999core}
J.~A. Russell and L.~F. Barrett, ``Core affect, prototypical emotional
  episodes, and other things called emotion: dissecting the elephant.''
  \emph{Journal of personality and social psychology}, vol.~76, no.~5, p. 805,
  1999.

\bibitem{russell1989affect}
J.~A. Russell, A.~Weiss, and G.~A. Mendelsohn, ``Affect grid: a single-item
  scale of pleasure and arousal.'' \emph{Journal of personality and social
  psychology}, vol.~57, no.~3, p. 493, 1989.

\bibitem{sakairi2013development}
Y.~Sakairi, K.~Nakatsuka, and T.~Shimizu, ``Development of the t wo-d
  imensional m ood s cale for self-monitoring and self-regulation of momentary
  mood states,'' \emph{Japanese Psychological Research}, vol.~55, no.~4, pp.
  338--349, 2013.

\bibitem{schwartz2016daily}
S.~Schwartz, S.~Schultz, A.~Reider, and E.~F. Saunders, ``Daily mood monitoring
  of symptoms using smartphones in bipolar disorder: a pilot study assessing
  the feasibility of ecological momentary assessment,'' \emph{Journal of
  Affective Disorders}, vol. 191, pp. 88--93, 2016.

\bibitem{senette2018persuasive}
C.~Senette, M.~C. Buzzi, M.~T. Paratore, and A.~Trujillo, ``Persuasive design
  of a mobile coaching app to encourage a healthy lifestyle during menopause,''
  in \emph{proceedings of the 17th International Conference on Mobile and
  Ubiquitous Multimedia}, 2018, pp. 47--58.

\bibitem{shiffman2008ecological}
S.~Shiffman, A.~A. Stone, and M.~R. Hufford, ``Ecological momentary
  assessment,'' \emph{Annu. Rev. Clin. Psychol.}, vol.~4, pp. 1--32, 2008.

\bibitem{spruijt2014dynamic}
D.~Spruijt-Metz and W.~Nilsen, ``Dynamic models of behavior for just-in-time
  adaptive interventions,'' \emph{IEEE Pervasive Computing}, vol.~13, no.~3,
  pp. 13--17, 2014.

\bibitem{trull2009using}
T.~J. Trull and U.~W. Ebner-Priemer, ``Using experience sampling
  methods/ecological momentary assessment (esm/ema) in clinical assessment and
  clinical research: introduction to the special section.'' 2009.

\bibitem{van2017experience}
N.~Van~Berkel, D.~Ferreira, and V.~Kostakos, ``The experience sampling method
  on mobile devices,'' \emph{ACM Computing Surveys (CSUR)}, vol.~50, no.~6, pp.
  1--40, 2017.

\bibitem{wallbaum2016comparison}
T.~Wallbaum, W.~Heuten, and S.~Boll, ``Comparison of in-situ mood input methods
  on mobile devices,'' in \emph{Proceedings of the 15th International
  Conference on Mobile and Ubiquitous Multimedia}, 2016, pp. 123--127.

\bibitem{watson1988development}
D.~Watson, L.~A. Clark, and A.~Tellegen, ``Development and validation of brief
  measures of positive and negative affect: the panas scales.'' \emph{Journal
  of personality and social psychology}, vol.~54, no.~6, p. 1063, 1988.

\bibitem{wilcoxon1950some}
F.~Wilcoxon, ``Some rapid approximate statistical procedures,'' \emph{Annals of
  the New York Academy of Sciences}, vol.~52, no.~6, pp. 808--814, 1950.

\end{thebibliography}
\balance
\end{document}